\documentclass[twocolumn,prb]{revtex4}

\usepackage{graphicx}

\def\be{\begin{equation}}
\def\ee{\end{equation}}
\def\ba{\begin{eqnarray}}
\def\ea{\end{eqnarray}}

\def\C60{A$_x$C$_{60}$}

\def\Y248{YBa$_2$Cu$_4$O$_8$}

\begin{document}

\title{Spectral function of a Luttinger liquid coupled to phonons
and angle-resolved photoemission measurements in the cuprate superconductors}

\author{Ian P. Bindloss and Steven A. Kivelson}
\affiliation{Department of Physics, University of California,  
Los Angeles, California 90095--1547, USA}
\date{\today}

\begin{abstract}

We compute the finite-temperature single-particle
spectral function of a one-dimensional Luttinger liquid coupled to
an optical phonon band.  The calculation is performed exactly for the 
case in which electron-phonon coupling
is purely forward scattering.  We
extend the results to include backward scattering with a renormalization group treatment.
The dispersion contains a change in velocity at
the phonon energy, qualitatively similar to the case of electron-phonon
coupling in a Fermi liquid.  If the backward scattering
part of the
the electron-phonon interaction
is not too strong compared to the forward scattering part,
coupling to phonons also produces a pronounced peak
in the spectral function at low energies.
The calculated spectral function is remarkably similar to the angle-resolved photoemission 
spectra of the high-temperature superconductors, including the
apparent presence of ``nodal quasiparticles,'' the presence of
a ``kink'' in the dispersion, and the non-Fermi-liquid
frequency and temperature dependencies. 
Although a microscopic justification
has not been established for treating the
electronic dynamics of the 
cuprates as quasi-one-dimensional,
at the very least we take the quality of the comparison as evidence
of the non-Fermi-liquid character of the measured spectra.

\end{abstract} 

\maketitle

\section{Introduction}
There is increasing, although controversial, experimental evidence 
that, in cuprate high-temperature superconductors,
a strong coupling between electrons and optical phonons  
produces observable features in the single-hole spectral function $A({\bf k},E)$ 
measured by angle-resolved photoemission spectroscopy (ARPES).\cite{shen1,shen2,lanzara}
In attempting to understand the implications of these measurements, the experiments have 
been previously analyzed in the context of the theory of the electron-phonon (el-ph) 
coupling in a Fermi liquid.
However, the observed spectral function exhibits
a manifestly non-Fermi-liquid frequency and temperature dependence,
\cite{anderson,drorprl,sachdev,vallaNFL} so the justification 
for this mode of analysis is not clear.

In the present paper, we compute and analyze the finite-temperature spectral function of a 
spinful,
one-dimensional (1D) Luttinger liquid (LL) coupled to a dispersionless optical phonon band 
(Einstein phonon).
Such a spectral function was computed previously in Ref. \onlinecite{meden},
but only at $T = 0$ and in the absence of both electron-electron (el-el) interactions and el-ph
backscattering.
The LL is a quantum critical point, so the spectral function is a scaling function,
and should be computed at finite temperature; the zero-temperature spectral function
reveals only the high-frequency behavior of the scaling function.
It is also essential to include el-el interactions, as in real materials they are typically
much stronger than the el-ph interactions.  Moreover, strong el-el interactions
make qualitative changes to the ``appearance'' of the spectral function.
The effects of phonon-assisted backward scattering are also important to consider,
as we have done below.

Our motivation for this study is threefold.  (1) 
There are a host of interesting quasi-1D materials,\cite{bourbonnais,nanotubes}
some of which are amenable to ARPES studies,\cite{gweon,allen,GweonAllen} to which this analysis 
may
be directly applicable.  (2)  The LL is the theoretically best
understood example of a non Fermi liquid--just as many aspects of the
Fermi liquid state are robust, independent of  details of material and
even dimensionality, we may hope that some features of non-Fermi-liquids
are similarly generic, at least within classes of non-Fermi-liquids. 
In this case, the LL may serve as a paradigmatic model for a broader
class of systems.
(3)  It may be the case that the
quasi-2D cuprates have significant, self-organized ``stripe''
structures\cite{stripereview} which render them locally quasi-1D, in 
which case it may be possible to directly compare the results
obtained here with experiments \cite{drorprl} in the cuprates and other highly
correlated materials.\cite{footnotestripes}

Let us start with a qualitative summary of the
solution of the el-ph problem in a Fermi liquid\cite{marsiglio}
at zero temperature.  Consider the 
case of an optical phonon with frequency $\omega_0$ and dimensionless
el-ph coupling $\lambda^\prime$, which is not too large.
For $|E| > \omega_0$, the effect of the el-ph coupling on  $A({\bf k},E)$
is an $E$-independent
broadening of the quasiparticle peak.
For $|E| \ll \omega_0$, the phonons 
can be integrated out to produce new effective interactions in the Fermi liquid:  the 
largest effect is a renormalization of the Fermi velocity, $v_F \to v_F^* = v_F / (1 
+ \lambda^\prime) < v_F$, while the most dramatic effect is the weak effective attraction 
produced 
between low-energy quasiparticles, which can lead to a superconducting instability of the
Fermi liquid state.

We have computed the single-hole spectral function for
a LL coupled to optical phonons for the case when the el-ph coupling does not
produce a spin gap.
We find that the same
{\it words} describe the effects of the el-ph coupling as in the
Fermi liquid case, but the results {\it look} quite different because the unperturbed LL
spectral function is not a simple Lorentzian. Specifically, if we let $\{g\}$ represent the set 
of 
coupling constants which define the LL (i.e. the charge and spin 
velocities $v_c$ and $v_s$, and the corresponding Luttinger exponents $K_c$ and 
$K_s$), then our result can be summarized by

\be
A(k,E) \sim \left\{\begin{array}{ll}
           A_{LL}(k,E; \{g\}) + \cdots  & \;    {\rm for }  \; |E| \gg \omega_0 , \\
	       A_{LL}(k,E; \{g^*\})     &     \;     {\rm for } \; |E| \ll \omega_0 ,
\end{array}
\right.
\label{approx}
\ee
and there is a smooth crossover between the two limits when $|E| \sim
\omega_0$.
Here, $A_{LL}$ is the spectral function of the pure LL
(i.e. in the absence of el-ph coupling),
the ellipsis represents perturbative  
corrections to the spectral function which can be ignored 
as long as the el-ph coupling is not too big or if $|E|$ is high enough,
and $\{g^*\}$ are renormalized
coupling constants obtained by integrating out the phonons--we give explicit expressions for
these renormalized couplings below.

The principal results of the present paper are Eqs. (\ref{approx}) and (\ref{main_result}).
The latter is an analytic expression for the space-time spectral function
that interpolates between the two limits in Eq. (\ref{approx}), and is shown
to be very nearly exact in the exactly solvable case of forward scattering only. 
We later generalize it for el-ph couplings that include
backscattering interactions.
Plots of $A(k,E)$ computed from Eq. (\ref{main_result})
are shown in the figures.

There are several general features of $A(k,E)$ that are worth noting.  
(1) As has been previously emphasized, \cite{drorprl} the spectral weight of the LL
is concentrated in a roughly triangular region of the $E$-$k$ plane
(see Fig. \ref{contour}), reflecting the 
fractionalized character of the elementary excitations, in contrast to the Fermi liquid case 
in which the spectral weight is concentrated along the line $E=v_F(k-k_F)$, reflecting the 
quasiparticle dispersion.  
(2)  There is a renormalization of the charge 
(holon) velocity produced by the el-ph coupling, such that $v_c^* < v_c$, 
which is analogous to the renormalization of $v_F$ in the Fermi liquid.  This
produces a ``kink'' in the spectrum, as shown in Fig. \ref{kink}.  
(3)  The $E$ dependence of $A({\bf k},E)$ at fixed ${\bf k}$ is referred to in the ARPES
literature as the energy distribution curve (EDC), while the ${\bf k}$ dependence at fixed $E$
is referred to as the momentum distribution curve (MDC); 
the extent to which there is a quasiparticle-like peak in the EDC of a LL is strongly dependent
on the value of
$K_c$.  For weakly interacting electrons, 
$K_c\approx 1$, and the EDC exhibits a peak, although this peak contains
a power-law tail indicating the absence of fermionic quasiparticles. 
For strong repulsive interactions that
are sufficiently long range, $K_c \ll 1$, in which case the EDC is extremely
broad, and can even fail to exhibit any well defined peak near the Fermi energy.
In contrast, the structure of the MDC is much less
variable,\cite{drorprl} and remains peaked even for small $K_c$.
The presence of extremely broad EDCs
at the same time that the MDCs are narrow
is a manifestly non-Fermi-liquid feature and a dramatic signature
of the LL.
This feature is illustrated in Fig. \ref{experiment_edc_vs_mdc}.

For a model with purely forward scattering el-ph interactions,
and for more general couplings as long
as the el-ph backscattering is not too strong,
at low energies $K_c$ is increased
by the presence of el-ph interactions, i.e. $K_c^* > K_c$.  For the physically
relevant case when the bare $K_c < 1$, this means that features that are situated within
$\omega_0$ of the Fermi energy are
made sharper (more peaked) by the coupling to phonons.
Therefore, for $K_c \ll 1$, the result has the following similarity with
the Fermi liquid case: features in $A(k,E)$ appear broader
for binding energies above $\omega_0$ than for binding energies below.
In the
Fermi liquid case, this comes from a phonon-induced broadening at high energies, but in the
LL it comes from a phonon-induced narrowing at low energies.\cite{footnotebroadening}

\begin{figure}
\includegraphics[width=0.39\textwidth]{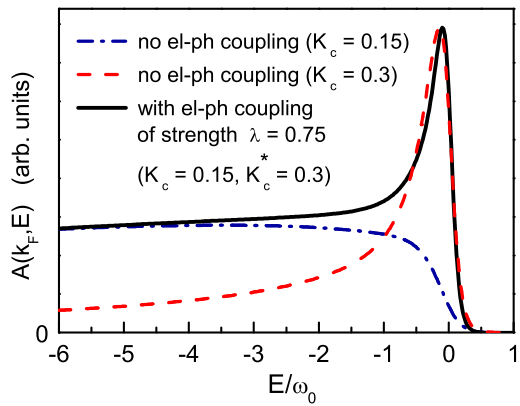}
\caption{\label{figure1}
Comparison of the single-hole spectral functions
at $k = k_F$ for
a LL coupled to optical phonons (solid line), 
and a LL in absence of phonon coupling (dashed and dash-dotted lines).
For the solid line, the el-ph coupling strength is $\lambda = 0.75$
($v_c/v_c^* = K_c^*/K_c = 2)$.
For all curves, $v_c/v_s = 4$ and $\omega_0/T = 10$.
See the plot for the values of $K_c$.
Normalizations are chosen for graphical clarity.
}
\end{figure}

A plot of the EDC at $k = k_F$ is shown in Fig. \ref{figure1}.
The solid line shows $A(k_F,E)$ for the case $K_c = 0.15$, in the presence of
forward scattering el-ph coupling, which produces, at low energies,
a renormalized $K_c^* = 0.3$.
The dashed line shows $A(k_F,E)$ for the pure LL, but with $K_c = 0.3$,
while the dash-dotted line shows it for the pure LL with 
$K_c = 0.15$.  Clearly, for the LL coupled to phonons,
the renormalized $K_c$ governs the properties
at low binding energies, while the bare $K_c$ dictates the behavior at 
large binding energies.

In Sec. \ref{exact_model}, we investigate the exactly solvable model
(forward scattering only).
The results are extended to general couplings using 
a renormalization group (RG) treatment in Sec. \ref{general}.
In Secs. \ref{experiment} and \ref{conclusions} we speculate on the relevance
of these results to real materials, especially the high-temperature superconductors.
Appendix \ref{App} contains a derivation of the spectral function of the forward
scattering only model.  There we also present the exact result for
the frequency- and momentum-dependent conductivity of this model.
The optical conductivity is unchanged by the el-ph forward scattering,
even though this interaction has dramatic effects on the spectral function. 
Appendix \ref{AppB} contains technical details for the results presented in Sec. \ref{general}.

\section{An exactly solvable model}
\label{exact_model}

The model is defined by the Hamiltonian density
\be
{\cal H} = {\cal H}_{\rm LL}+{\cal H}_{\rm ph}+{\cal H}_{\rm el-ph} \:.
\ee
Here the purely electronic part of the Hamiltonian
\be
{\cal H}_{\rm LL} = \sum_{\alpha = c,s} \frac {v_{\alpha}} 2 \left[ K_{\alpha}\Pi_{\alpha}^2
+ \frac {(\partial_x\phi_{\alpha})^2} {K_{\alpha}}\right ]
\label{H_LL}
\ee
is the famous spin-charge-separated Tomonaga-Luttinger liquid model 
of the interacting one-dimensional electron gas (1DEG) at incommensurate filling.
It is expressed
via bosonization, in terms of
bosonic charge ($\alpha = c$) and spin ($\alpha = s$) fields $\phi_{\alpha}$, and their
canonically conjugate momenta $\Pi_{\alpha}$.
Repulsive interactions usually renormalize the Luttinger parameter $K_c$
below its noninteracting value of 1 such that
$0 < K_c < 1$, and renormalize the velocities such that the $v_s < v_F < v_c$.
We will assume the system is spin-rotation invariant, which dictates that $K_s = 1$.
Expressions for the $v_\alpha$'s and $K_\alpha$'s
in terms of microscopic short-range interaction parameters
and reviews on the technique of bosonization can be found in
many places in the literature.\cite{emery,fradkinbook,voit,tsvelik,giamarchi}
Bosonization allows the fermionic fields to be expressed directly
in terms of the bosonic fields:
\be
\Psi_{\eta,\sigma} = \frac{e^{i \eta k_F x}} {\sqrt{2 \pi a}}
\exp\left\{ i\sqrt{\frac{\pi} {2}}\left[\eta(\phi_c+\sigma\phi_s) +
\theta_c +\sigma\theta_s\right]\right\} ,
\ee
where $\Psi_{\eta,\sigma}$ is the right- or left-moving 
fermionic destruction field ($\eta = \pm 1$, respectively)
for spin up or down ($\sigma= \pm 1$, respectively).
Here $\theta_{\alpha}(x) = -\int_{-\infty}^x dx^{\prime} \Pi_{\alpha}(x^{\prime})$ labels the 
dual
bosonic field, and $a$ is a short-distance cutoff corresponding to the
lattice parameter.  As is well known, the Hamiltonian
in Eq. (\ref{H_LL}) describes
a line of fixed points for interacting electrons, and so captures
the essential low-energy physics of a large class of physical systems. 

The purely vibrational part of the Hamiltonian
\be
{\cal H}_{\rm ph} = \left[ P^2 + \omega_0^2 u^2\right]/2M
\ee
describes an Einstein oscillator.  Here $u$ and $P$ are the phonon field and its canonical
conjugate momentum,
$M$ is the ion mass, and again $\omega_0$ is the optical phonon frequency.
Note that we work with units such that Boltzmann's and Planck's constants are
$k_B=\hbar=1$.

In the following section, we will consider a general form of the el-ph
coupling, ${\cal H}_{\rm el-ph}$, but for the purposes of the present section, we consider
forward scattering interactions only,
\be
{\cal H}_{\rm el-ph}= \alpha_2 \ u\  \hat \rho = \alpha_2 \sqrt{2/\pi} \ u\  (\partial_x\phi_c),
\label{H_el_ph}
\ee
where $\alpha_2$ is the el-ph
coupling parameter, $\hat\rho$ is the long-wavelength component of the charge density,
and the second equality makes use of the standard bosonization expression for $\hat \rho$. 
viet
The forward scattering model is exactly solvable
since in its bosonized form it is quadratic in the fields.  In Appendix \ref{App} we
compute the renormalized couplings that define the $\{g^*\}$ for this model.
The exact result is
\be
v_c^* = v_c \sqrt{1 - \lambda} \  ,  \ \ K_c^* = \frac{K_c}{\sqrt{1 - \lambda}} \ , 
\label{vc_star_Kc_star}
\ee
where
\be
\lambda = \frac{2 K_c \alpha_2^2} {\pi v_c M \omega_0^2} \: ,
\label{lambda}
\ee
and the spin couplings are unrenormalized.
Note that $\lambda$ depends on both the el-ph and el-el interactions.
Also note that $v_c/v_c^* = K_c^*/K_c \ge 1$. At low energies, $v_c$
is reduced due to ``phonon drag,'' while $K_c$ is increased due
to an attractive interaction mediated by phonons.
Since this model possesses
an instability at $\lambda=1$, where the charge velocity goes to zero,
$\lambda$ is restricted to the range $0 \le \lambda < 1$.  The analog
of this instability for the case of coupling to {\it acoustic} phonons has been studied
previously.\cite{MartinLossPRB}

In Appendix \ref{App} we derive an exact expression for the spectral function.  This quantity
is most easily expressed in position space, but even then
involves a momentum integral in
the exponent that
cannot be performed analytically.
The integral can be evaluated numerically,
and fortunately we are able to derive
a simple analytic expression that accurately approximates it,
as shown in Appendix \ref{App}
and Fig. \ref{exact_vs_approx}.
We therefore use this analytic approximation in subsequent
calculations.  

Specifically, our analytic approximation for
the single-hole Green's function
$G_\eta(x,t;\lambda) = \left<\Psi_{\eta,\sigma}^{\dagger}(x,t)\Psi_{\eta,\sigma}(0,0)\right>$ 
is 
\be
G_\eta(x,t;\lambda) \approx
G_\eta(x,t;0) \frac {g_\eta(x,t;v_c^*,K_c^*,v_c/\omega_0)} {g_\eta(x,t;v_c,K_c,v_c/\omega_0)} \: 
,
\label{main_result}
\ee
where the exact
Green's function in the absence of phonon
coupling $G_\eta(x,t;0)$ at temperature $T = 1/\beta$ is\cite{pureLLspectralfunction}
\be
G_\eta(x,t;0) = \frac{e^{-i \eta k_F x}} {2 \pi a}\prod_{\alpha=c,s} 
{g_\eta(x,t;v_\alpha,K_\alpha,a}) \:,
\ee
and we have defined the function
\ba
\nonumber &&g_\eta(x,t;v,K,a)\\
&&= \prod_{j=\pm 1}\left\{ \frac{i v \beta}{\pi a} \textnormal {sinh}\left[ \frac{\pi(vt + j \eta 
x - ia)}{v \beta}\right]\right\}^{-(K-j)^2/8K} . \; \; \;
\label{g}
\ea
Equation (\ref{main_result}) is the central result of this paper. 
The spectral function measured by ARPES
is accurately given by its Fourier transform.

\begin{figure}
\includegraphics[width=0.46\textwidth]{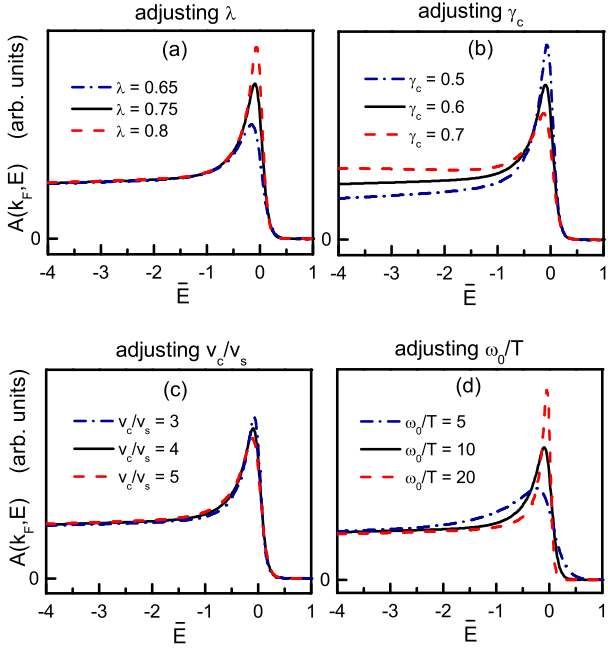}
\caption{\label{parameters} Dependence of the EDC at $k = k_F$
on $\lambda$, $\gamma_c$, $v_c/v_s$, and $\omega_0/T$
[(a), (b), (c), and (d) respectively].  In each panel,
only one parameter is varied, and the rest are held fixed.  Unless
otherwise labeled, $\lambda = 0.75$, $\gamma_c = 0.6$, $v_c/v_s = 4$,
and $\omega_0/T = 10$.  For example, for the curves in (a), $\lambda$ has the value
labeled in the plot legend, while $\gamma_c = 0.6$, $v_c/v_s = 4$, and $\omega_0/T = 10$.
In (a), the values
of $v_c/v_c^* = K_c^*/K_c$ are 1.69, 2, 2.24 for
$\lambda$ equal to 0.65, 0.75, 0.8, respectively.
In (b), $K_c$ is 0.172, 0.150, 0.134
for $\gamma_c$ equal to 0.5, 0.6, 0.7, respectively. 
Normalizations are chosen for graphical clarity
(the apparent variation of the total spectral weight
is an artifact). 
}
\end{figure}

Specifically, the single-hole
spectral function for right-moving fermions is\cite{footnotesinglehole} 
\be
A(k,E) = \int_{-\infty}^{\infty}dx \int_{-\infty}^{\infty}dt \: e^{i(kx - E t)} \, 
G_1(x,t;\lambda) \: .
\label{fourier}
\ee
By combining this with Eq. (\ref{main_result}), one obtains
\ba
\nonumber A(k, E) &\approx& \frac{{\bar \beta}^2}{\pi {\bar a} \omega_0} \int_{-\infty}^\infty 
d{\tilde x} \int_0^\infty d{\tilde t} \: \chi({\tilde x},{\tilde t}) \; \; \; \; \; \\
&\times& \cos\left[{\bar \beta}({\bar k}{\tilde x}-{\bar E}{\tilde t}) - \Theta({\tilde 
x},{\tilde t})\right] , \; \; \; 
\label{first}
\ea
where we defined the dimensionless quantities
${\bar \beta} = \omega_0 \beta /\pi$,  
${\bar a} = a \omega_0 /v_c$,
\be
{\bar k} = v_c(k-k_F)/\omega_0 , \; \; \; {\bar E} = E/\omega_0 ,
\ee
and the functions
\be
\chi({\tilde x},{\tilde t}) = \prod_{i=1}^4 \prod_{j=\pm 1}\left[ 
\frac{(a_i/v_i)^2}{\sinh^2({\tilde t}+j{\tilde x}/v_i)+\sin^2(a_i/v_i)} \right]^{\gamma_{ij}/2}
\ee
and
\be
\Theta({\tilde x},{\tilde t}) = \sum_{i=1}^4 \sum_{j=\pm 1} \, \gamma_{ij} \, 
\arctan\left[\frac{\tanh({\tilde t}+j{\tilde x}/v_i)}{\tan(a_i/v_i)}\right],
\ee
with
\ba
&&\gamma_{ij} = (1-2\delta_{i3})\frac{(K_i - j)^2}{8K_i} , \label{gamma} \; \; \; \; \; \\
&&K_1 = 1 , \; \; K_2 = K_3 = K_c , \; \; K_4 = K_c^* , \; \; \; \; \; \\
&&v_1 = v_s/v_c , \; \; v_2 = v_3 = 1 , \; \; v_4 = v_c^*/v_c , \; \; \; \; \; \\
&&a_1 = a_2 = {\bar a}/{\bar \beta} , \; \; a_3 = a_4 = 1/{\bar \beta} \, . \; \; \; \; \; \label 
{last}
\ea
Above, ${\tilde x}$ and ${\tilde t}$ are dimensionless dummy variables of integration,
$\delta_{i3}$ is the Kronecker delta, and ${\bar a} \ll 1$ is a dimensionless cutoff. 
The spectral function plots
were obtained by performing the double
integration in Eq. (\ref{first}) numerically.
Henceforth, we adopt the notation
\be
\gamma_c \equiv (K_c + K_c^{-1} - 2)/8 = \gamma_{21} \, ,
\ee
which vanishes in the absence of el-el interactions.

$A(k,E)$ depends only on ${\bar k}$, ${\bar E}$,
$\omega_0/T$, and on the three
dimensionless parameters $\lambda$, $\gamma_c$, and $v_c/v_s$.\cite{footnoteunits}   
In order to provide the reader with a qualitative understanding
of how the spectral function depends on these parameters, in Fig. \ref{parameters}
we show how the EDC at $k = k_F$ changes when one is varied
with the rest held fixed.  Henceforth,
we either chose representative parameters for the figures,
or, for cases in which the theory is compared to experimental data,
we fit the parameters.
Not surprisingly, for cases in which we have fitted the parameters,
they turn out to be somewhat material dependent. 

\begin{figure}
\includegraphics[width=0.43\textwidth]{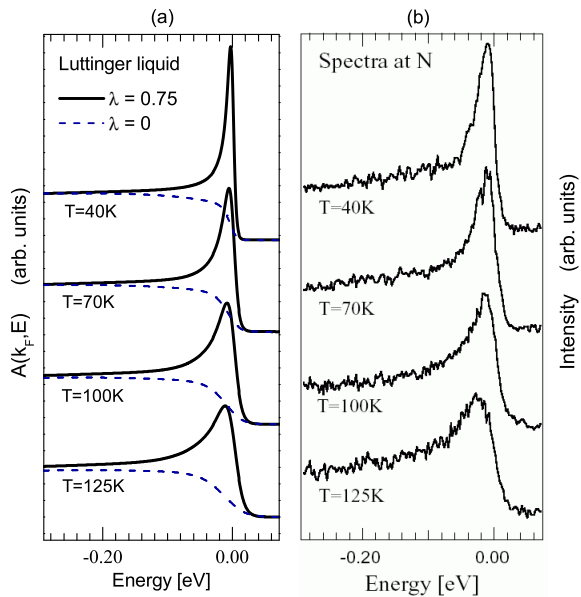}
\caption{\label{temperature} Comparison of the temperature dependence of the
EDCs at $k = k_F$ for (a) a LL coupled to phonons
and (b) ARPES data of optimally doped Bi2212
($T_c = 89 \; \rm K$)
at the Fermi surface crossing along 
${\bf k} = (0,0)$ to ($\pi$,$\pi$)
(the nodal direction) (Ref. \protect\onlinecite{Campuzano}).
In (a), $\omega_0 = 70 \; {\rm meV}$, $\gamma_c = 0.6$, and $v_c/v_s = 4$.
The solid line is for $\lambda = 0.75$, and the dashed
line is the result for the pure LL.
}
\end{figure}

Figure \ref{temperature}(a) exhibits the temperature dependence of the EDC at 
$k = k_F$.
The spectral weight of the low energy peak is reduced
by increasing $T$, while at the same time the width of the peak 
increases proportional to $T$, due to the quantum critical nature of the LL. 
The value $\gamma_c = 0.6$ ($K_c \approx 0.15$) used
here indicates strong el-el repulsion.

\begin{figure}
\includegraphics[width=0.25\textwidth]{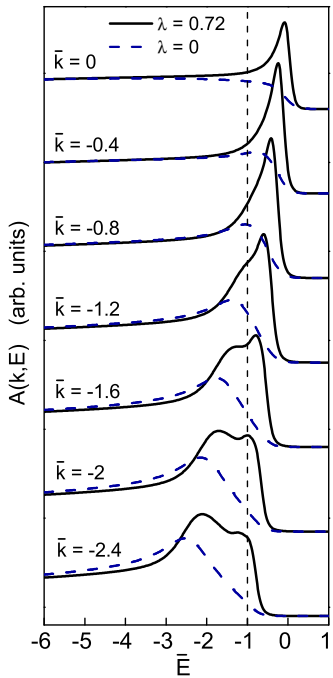}
\caption{\label{edc_theory_only} EDCs with (solid lines) and without (dashed lines)
forward scattering coupling to phonons of strength $\lambda = 0.72$
($v_c/v_c^* = K_c^*/K_c = 1.9$),
with $\gamma_c = 0.6$, $v_c/v_s = 3$, and $\omega_0/T = 10$.  The
vertical dashed line is drawn at the phonon energy.  
The notation is ${\bar k}= v_c (k - k_F)/\omega_0$
and ${\bar E} = E/\omega_0$,
where $E$ is measured with respect to the Fermi energy.
The curves have vertical offsets proportional to ${\bar k}$.
}
\end{figure}

In Fig. \ref{edc_theory_only} we show the dependence of the EDCs
on ${\bar k}$, with and without forward
scattering coupling to phonons, for $\omega_0/T = 10$
and $\gamma_c = 0.6$.  Note the ``double-peak'' structure
present near the phonon energy for moderate values of ${\bar k}$.

\begin{figure}
\includegraphics[width=0.44\textwidth]{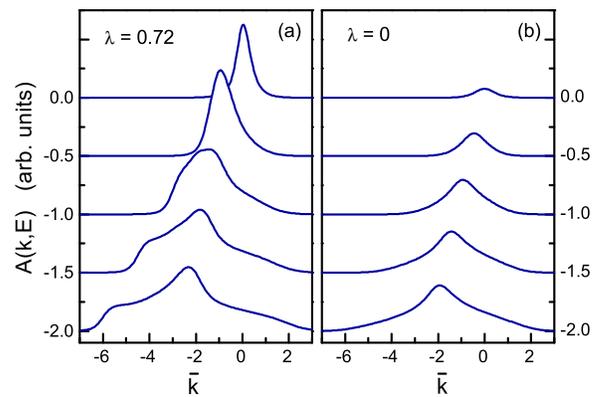}
\caption{\label{MDCs_high_T} MDCs (a) with and (b) without
coupling to phonons of strength $\lambda = 0.72$
for $\omega_0/T = 10$.  All
parameters are the same as Fig. \ref{edc_theory_only}.
The curves are offset by ${\bar E}$.
}
\end{figure}

\begin{figure}
\includegraphics[width=0.44\textwidth]{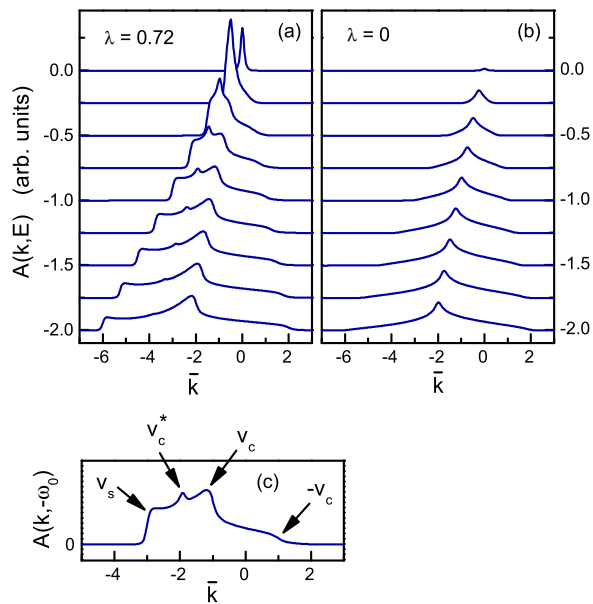}
\caption{\label{MDCs} MDCs (a) with and (b) without
coupling to phonons for
$\omega_0/T = 40$ and the same
interaction strengths as Fig. \ref{edc_theory_only}.
(c) shows an enlargement of
the ${\bar E} = -1$ curve from (a), and labels the velocities
at which various features disperse.  In (a) and (b)
the curves are offset by ${\bar E}$.
}
\end{figure}

Figure \ref{MDCs_high_T} shows MDCs at various values of ${\bar E}$ for the same parameters
as Fig. \ref{edc_theory_only}.  
A similar plot is shown in Fig. \ref{MDCs}
at a lower temperature ($\omega_0/T = 40$).
Because of the lower temperature,  
here one can resolve {\it three} local maxima
for moderate values of ${\bar E}$.  As shown
in Fig. \ref{MDCs}(c), which is an enlargement of the ${\bar E} = -1$ curve,
the center local maximum disperses at the renormalized
charge velocity $v_c^*$.
Since this peak is by far the dominant one for $|{\bar E}| \ll 1$,
the low-energy dispersion is characterized
by what appears to be a single peak dispersing with velocity
$v_c^*$.
At higher $|{\bar E}|$, the $v_c^*$ peak disappears.
If the temperature is increased sufficiently, the
three local maxima can no longer be resolved 
and appear as one peak, as seen in Fig. \ref{MDCs_high_T}(a).
Also note that phonon coupling
creates larger spectral weight
at the spinon velocity $v_s$, due
to the higher effective $K_c$
(this feature disappears at high binding energies).
 
\begin{figure}
\includegraphics[width=0.49\textwidth]{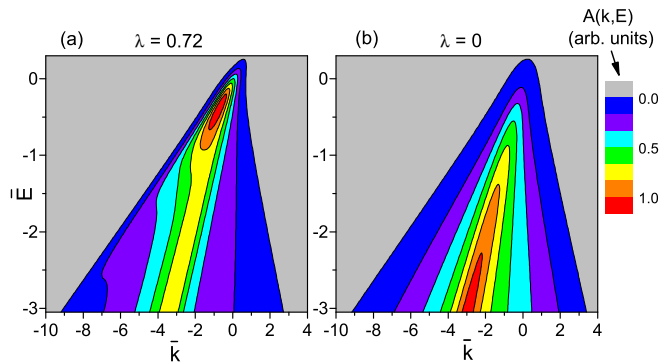}
\caption{\label{contour} Contour plot for $\omega_0/T = 10$
of the single-hole spectral function
(a) with and (b) without
forward scattering to phonons of strength $\lambda = 0.72$.  All parameters
are the same as in Fig. \ref{edc_theory_only}.
}
\end{figure}

In Fig. \ref{contour} we present contour plots in the ${\bar E}$-${\bar k}$ plane,
with and without phonon coupling, for the same parameters as Fig.
\ref{edc_theory_only}.  Note the pronounced peak (red spot)
at low binding energies in Fig. \ref{contour}(a),
due to the increase in the effective $K_c$.  The reduction
in charge velocity at low binding energies can also be seen here.

\begin{figure}
\includegraphics[width=0.23\textwidth]{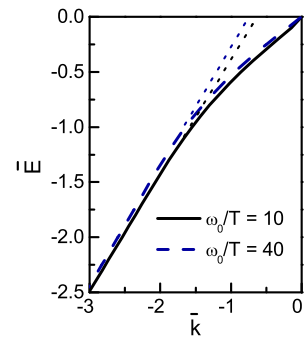}
\caption{\label{kink} Dispersion for $\lambda = 0.72$,
determined
by fitting MDCs to Lorentzians, for
$\omega_0/T = 10$ (solid line) and
$\omega_0/T = 40$ (dashed line).
Interaction strengths are the same as in Fig. \ref{edc_theory_only}.
The dotted lines are fits
to the high-$|{\bar E}|$ portion of the dispersions.
}
\end{figure}

Figure \ref{kink} shows the dispersion
in the ${\bar E}$-${\bar k}$ plane, determined by
fitting MDC curves to Lorentzian functions (the
same method used in the ARPES literature).
A change in the slope
occurs at ${\bar E} \approx -1$.  The ratio of the slope at $|{\bar E}| \ll 1$
to the slope at $|{\bar E}| \gg 1$ is approximately
$v_c^*/v_c = \sqrt{1 - \lambda}$.
Note that the dispersion, when determined by fitting to Lorentzians, is weakly $T$ dependent.

\section{General electron-phonon coupling}
\label{general}
In general, both forward and backward scattering (i.e. with momentum transfer near
$2k_F$) are possible.  Thus, in general, we should consider both processes:
\be
\label{general_el_ph}{\cal H}_{\rm el-ph} = u \left[\alpha_2  \ \hat \rho + \alpha_1 
\sum_{\sigma}\left(\Psi_{1,\sigma}^{\dagger}\Psi_{-1,\sigma} + {\rm H.c.}\right)\right].
\ee
In this case, because $\Psi$ is a nonlinear function of the bosonic fields, the problem is not
exactly solvable.  We therefore treat the backscattering term $\alpha_1$ with
a perturbative renormalization group scheme.  It is important to note that if $\alpha_1$ is 
sufficiently strong,
a gap opens up in the spin sector, and the system is a Luther-Emery liquid\cite{LEL} (LEL) 
instead
of a LL (see Appendix \ref{AppB}).  We refer the reader to Ref. \onlinecite{ian} for a detailed
study of the phase boundary separating the LL and LEL phases. 
Here we limit ourselves to the case in which the spectrum is gapless.

\begin{figure}
\includegraphics[width=0.31\textwidth]{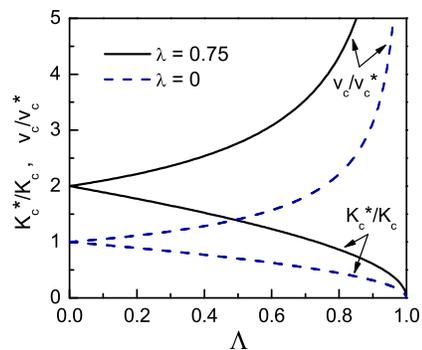}
\caption{\label{backscattering} Dependence of the magnitude of the kink ($v_c/v_c^*$)
and the degree to which the spectral function is peaked at low energies compared to high energies
($K_c^*/K_c$) on the effective backscattering el-ph interaction parameter $\Lambda$, for the case
$K_c = 0.15$.  The dashed lines are the result in the absence of forward scattering
el-ph interaction ($\lambda = 0$), while the solid lines are for $\lambda = 0.75$.
}
\end{figure}

It is convenient
to define the dimensionless el-ph couplings
\be
\label{barecouplings} \lambda_1 = \frac{\alpha_1^2} {\pi v_F M \omega_0^2} \: , \;\;\; \lambda_2 
= \frac{\alpha_2^2} {\pi v_F M \omega_0^2} \: .
\ee
In Appendix \ref{AppB} we derive the following relations,
which are generalizations of Eq. (\ref{vc_star_Kc_star}) to the case of nonzero $\lambda_1$:
\ba
\label{new_vc} v_c^* &=& v_c \sqrt{(1 - \Lambda)(1 - \lambda + K_c^2 \Lambda)} \: , \\
\label{new_Kc} K_c^* &=& K_c \sqrt{\frac{1 - \Lambda}{1 - \lambda + K_c^2 \Lambda}} \: ,
\ea
where we have introduced the effective backscattering el-ph parameter
\be
\label{cap_lambda} \Lambda = \frac{v_F}{2 K_c v_c} \lambda_1^* , \; \; 
\ee
which lies in the range $0 \le \Lambda < 1$.
Here $\lambda_1^*$ is the renormalized value of the bare el-ph backscattering parameter 
$\lambda_1$.
For the case in which the el-ph interaction is unretarded (when the Fermi energy $E_F < 
\omega_0$),
$\lambda_1$ remains unrenormalized ($\lambda_1^* = \lambda_1$).  However, for the physically 
interesting case $E_F > \omega_0$,
$\lambda_1^*$ depends on the ratio $E_F/\omega_0$ and on the strength of the el-el
interactions (see Appendix \ref{AppB} for an explicit relation).
As before, the forward scattering el-ph parameter is $\lambda = (2K_c v_F/v_c) \, \lambda_2$;
note that $\lambda_2$ contains no asterisk because it remains unrenormalized regardless of 
$E_F/\omega_0$.\cite{footnoteinteresting}

Equations (\ref{new_vc}) and (\ref{new_Kc})
are actually completely general, nonperturbative expressions.
However, our only way of relating the effective parameter $\Lambda$
to the microscopic parameter $\lambda_1$ is through Eq. (\ref{lambda_star}), which 
was obtained from one-loop RG, and
is therefore valid only when the el-el interactions are weak and $\lambda_1 \ll 1$.
But Eqs. (\ref{new_vc}) and (\ref{new_Kc}) remain valid even if the RG is
carried out to an infinite number of loops.

If fitting to particular ARPES data, the parameters 
$K_c$, $K_c^*$, and $v_c^*/v_c$ are easily obtained.
Then, $\lambda$ and $\Lambda$ can be determined
by inverting Eqs. (\ref{new_vc}) and (\ref{new_Kc}).
Therefore, for quasi-1D systems, ARPES
is an effective probe of the relative amounts of forward
and (renormalized) backward scattering el-ph interactions.

Note that for nonzero $\Lambda$, the
relation $K_c^*/K_c = v_c/v_c^*$ no longer holds; instead 
$K_c^*/K_c = (1 - \Lambda)v_c/v_c^*$.
For $K_c < 1$, the relation
$v_c^* < v_c$ holds regardless of $\Lambda$ and $\lambda$.
However, $K_c^* > K_c$ holds only if the ratio $\Lambda/\lambda$ is small enough. 
Specifically, if $\Lambda/\lambda > 1/(1 + K_c^2)$, then $K_c^* < K_c$, which means that
the low-energy spectral features
are no longer made more peaked due to phonons.
Figure \ref{backscattering} illustrates the result of increasing $\Lambda$ at fixed $\lambda$, 
for the case
$K_c = 0.15$.
The magnitude of the kink, given by $v_c/v_c^*$, is {\it increased} by turning
up $\Lambda$, while $K_c^*$ is reduced.  For all the spectral function plots in this
paper, we have set $\Lambda = 0$.
 
\section{Comparisons with ARPES experiments in the cuprates}
\label{experiment}
In this section, we wish to illustrate the extent to which the observed ARPES spectra in the
cuprate superconductors resemble those of a LL coupled to optical phonons.
At a gross level, the character of the
resemblance between the observed spectrum and a pure LL
was established in Ref. \onlinecite{drorprl}.  The strength of this analogy is
further supported \cite{GweonAllen} by direct comparison between the experimentally measured
$A(k,E)$ in a quasi-1D bronze and the cuprates.  However, as more and
better data have become available, it has become clear that there are features in the cuprate
data--especially the widely reported ``kinks'' in the dispersion of the MDC 
peaks\cite{shen1,shen2,shen3,CampuzanoKink,johnson}--that are qualitatively absent from the pure 
LL.  The pure LL
is also unable to reproduce the ``nodal quasiparticles''
seen in experiments while simultaneously producing broad features at high energy.
Here, therefore, we propose to
make a comparison between the measured spectral functions, and the spectral
function of a LL coupled to an optical phonon.\cite{footnotebackground}
Since the LL describes a gapless state,
we only compare our results to data taken in the
gapless ``nodal'' direction, which is defined as ${\bf k} = (0,0)$ to ($\pi$,$\pi$).
In other directions, the ARPES spectrum of the cuprates develops a gap below a certain
temperature, whereas the spectrum remains gapless in the nodal direction
even in the superconducting state.

In Fig. \ref{temperature} we have fitted the theoretical
EDCs to ARPES data\cite{vallaNFL}
in optimally doped ${\rm Bi_2 Sr_2 Ca Cu_2 O_{8+{\it \delta}}}$  
(Bi2212)\cite{Campuzano} at the Fermi surface crossing in the nodal direction,
shown at various temperatures, both above and below $T_c$.  The experimental temperature
dependence has been previously
interpreted as evidence for quantum critical behavior.
Since the LL is a quantum critical state, the resemblance
between theory and experiment in the present paper
supports this interpretation.

\begin{figure}
\includegraphics[width=0.49\textwidth]{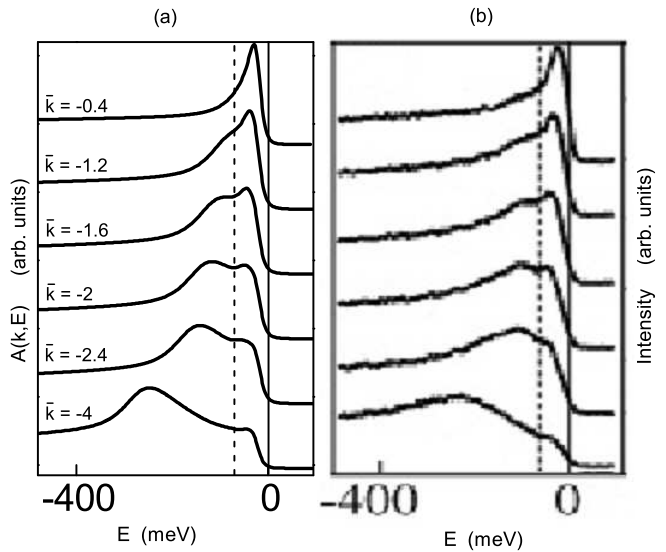}
\caption{\label{EDCs} Comparison of EDCs for (a) a LL coupled to phonons
and (b) ARPES data of slightly underdoped Bi2212 along the nodal direction ($T_c = 84 \; \rm K)$
at various momenta (Ref. \protect\onlinecite{shen1}).
For (a), $\lambda = 0.72$, $\gamma_c = 0.6$, $v_c/v_s = 3$, $\omega_0 = 70 \; {\rm meV}$, $T = 80 
\; \rm K$,
and the dashed line is drawn at $|E| = \omega_0$.  
}
\end{figure}

Figure \ref{EDCs} presents a fit to normal state EDCs for  
slightly underdoped\cite{shen1} Bi2212 along the nodal direction, at various momenta.
The theoretical curves contain a similar double-peak structure
as the experiment.  The line shapes of both the theory and experiment
are characterized by extremely long high energy tails.

\begin{figure}
\includegraphics[width=0.4\textwidth]{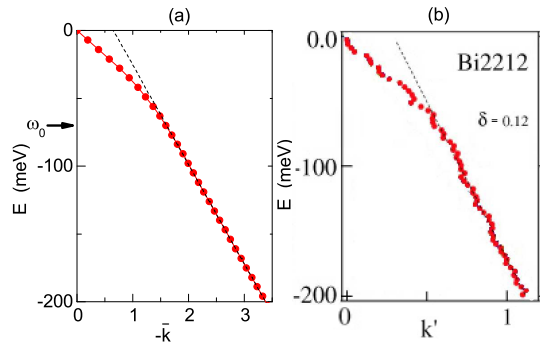}
\caption{\label{peaks} Comparison of the dispersion of (a) the LL coupled to phonons
with $\omega_0 = 70 \; {\rm meV}$
at $T = 30 \; \rm K$
to (b) the dispersion in Bi2212 ($T_c = 84 \; \rm K$) at $T = 30 \; \rm K$ in the nodal direction 
(Ref. \protect\onlinecite{shen1}).
For (a), all interaction strengths are the same as in Fig. \ref{EDCs}.
For both plots, the dispersion was obtained by fitting MDC curves to Lorentzians.
In (b), the authors define the rescaled momentum $k^\prime$,
by normalizing to 1 the
value of $k_F - k$ at $E = -170 \; {\rm meV}$. 
The dashed lines are guides to the eye.
}
\end{figure} 

\begin{figure}
\includegraphics[width=0.48\textwidth]{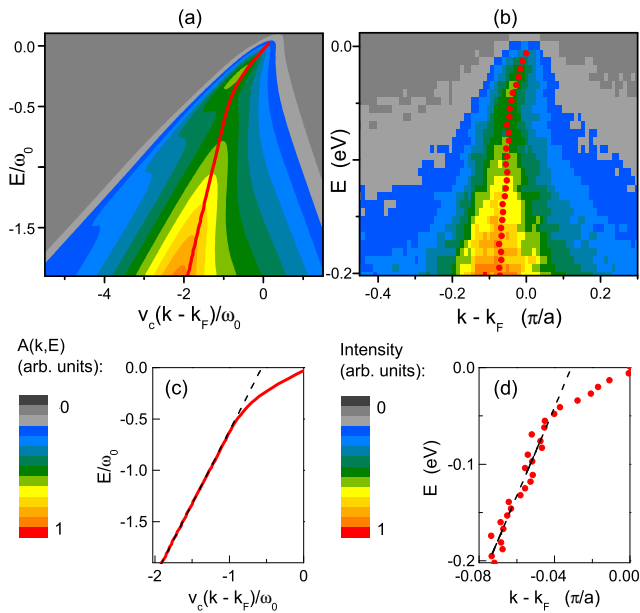}
\caption{\label{contour_experiment} Comparison of intensity plot of (a)
the single-hole spectral function of a LL coupled to phonons
with $\gamma_c = 1.5$, $\lambda = 0.93$,
$\omega_0/T = 20$, and $v_c/v_s = 3$ to
(b) the
ARPES spectrum of nonsuperconducting LSCO (3\% doping) at $T = 30 \; \rm K$ (Ref. 
\protect\onlinecite{LSCO}). 
In (a) and (b), the red line and points
are the dispersion, obtained by fitting MDCs to Lorentzians.
Panels (c) and (d) show
enlargements of the dispersions; the dashed lines are
fits to the high-binding-energy portion.
}
\end{figure}

Figure \ref{peaks}(a) shows the theoretical dispersion
for the same parameters as Fig. \ref{EDCs}(a), except with $T = 30 \; \rm K$,
determined by fitting the
theoretical MDCs to Lorentzians.  Fig. 
\ref{peaks}(b) shows the experimental dispersion for slightly underdoped\cite{shen1} Bi2212 in 
the nodal
direction at $T = 30 \; \rm K$, which the authors obtained
with the same fitting procedure.
The bizarre feature of the experimental data in which 
the dispersion at high binding energies does not extrapolate to the origin is also seen
in the theory.\cite{footnotekink}

For the theory plots,
the manner by which the dispersion is
extracted (least-squares fitting to Lorentzians),
yields
a weighted average of the {\it four}
velocities labeled in Fig. \ref{MDCs}(c). 
It is the
contribution from the velocity $-v_c$
that is responsible for the plotted high-energy dispersion
not extrapolating to the origin at $E = 0$ and $k = k_F$.
This is because,
when the dispersion is determined by fits to Lorentzians,
the presence of spectral weight at $k > k_F$
[near the right side of the triangle in Fig. \ref{contour_experiment}(a)]
``pulls'' the high-binding-energy
dispersion to higher velocities.

\begin{figure}
\includegraphics[width=0.34\textwidth]{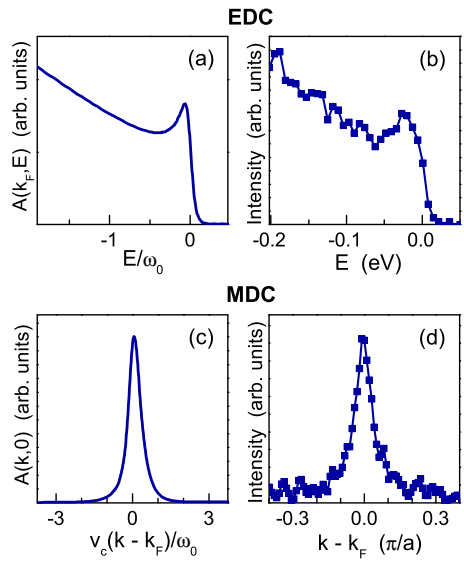}
\caption{\label{experiment_edc_vs_mdc} EDC at $k = k_F$ for (a) LL coupled
to phonons and (b) underdoped nonsuperconducting LSCO (Ref. \protect\onlinecite{LSCO}) (nodal 
direction); and
MDC at $E = 0$ for (c) LL coupled
to phonons and (d) underdoped LSCO (nodal direction).
Panels (a) and (c) are slices from the plot
in Fig. \ref{contour_experiment}(a); 
(b) and (d) are slices from the data in
Fig. \ref{contour_experiment}(b).
}
\end{figure}
 
Figure \ref{contour_experiment}(a) shows
a contour plot of the theoretical spectral function
in the ${\bar E}$-${\bar k}$ plane 
and compares it to ARPES data\cite{LSCO} for underdoped, nonsuperconducting
${\rm La_{2-{\it x}} Sr_{\it x} Cu O_4}$ (LSCO) with $x = 0.03$ [Fig. 
\ref{contour_experiment}(b)].
The value of $\lambda$ was chosen to give
$v_c/v_c^* = K_c^*/K_c = 3.8$, which is the same as
the ratio of the high to low binding energy velocities in the experimental plot.
In Figs. \ref{contour_experiment}(c) and \ref{contour_experiment}(d) we show the dispersions,
obtained in the same manner as in Fig. \ref{peaks}.
The EDCs and MDCs
from these plots, at the Fermi momentum and Fermi energy, respectively, are shown explicitly
in Fig. \ref{experiment_edc_vs_mdc}.  For both the theory and experiment,
the contrast between the
sharpness of the MDCs and the breadth of the EDCs is dramatic.

Note that the large values of $\gamma_c$ and large
values of $\lambda$
used for all the theoretical plots above indicate the presence
of very strong el-el and el-ph interactions, with an 
effective el-ph interaction
that is peaked in the forward scattering direction (since we used
$\Lambda = 0$).

In Refs. \onlinecite{Verga} and \onlinecite{Schachinger},
an effort is made to
fit the experimental dispersion
of the cuprates to the conventional theory of el-ph coupling in a Fermi liquid.
Both papers report that getting a good fit to the experimental kink
requires coupling between electrons and a broad spectrum of phonon modes
that extends all the way up to two or three hundred meV.
As the authors point out,
this is unphysical, since from neutron scattering, the highest-energy phonon mode
is less than 100 meV in all of the cuprates.  This suggests
that either phonons are not responsible for the kink, or if they are,
that the correct theory is very different from the conventional
Fermi liquid one.  In the present paper,
the phonon density of states that couples to electrons is a single
$\delta$ function at $\omega_0$ (Einstein phonon).
But even treating the phonon density of states as adjustable,
we suspect that a Fermi liquid treatment would not be able to fit the
frequency and temperature dependence of the spectral function, nor the
linear dependence of the MDC width on energy.\cite{vallaNFL} 
 
\section{Conclusions}
\label{conclusions}

We have analyzed the effect of the el-ph coupling on the single-particle
spectral function of the theoretically best understood (and exactly solvable) non-Fermi-liquid,
the LL.  Since, by definition, a non-Fermi-liquid is a state in which the elementary excitations
are not simply dressed electrons or holes, $A({\bf k},E)$ should be the measurable
quantity in which non-Fermi-liquid effects are most dramatic.  Thus, it is important to have a
clear idea of which features of this function best distinguish a Fermi liquid from a
non-Fermi-liquid.  It has been argued previously \cite{drorprl} that the most dramatic
signature of a LL is the appearance of extremely broad tails in the EDC, although peaks
in the MDC remain relatively sharp.  (This signature is particularly dramatic
when the el-el
interactions are strong; while the same distinction applies {\it in principle} for weak
interactions, {\it in practice} the LL is harder to distinguish from a Fermi liquid
when the couplings are weak.)

We have shown that some of the
gross characteristic features of el-ph coupling in a Fermi liquid--an apparent kink
in the dispersion relations and EDCs that are more peaked at $|E| < \omega_0$
compared to $|E| > \omega_0$--are also present in a LL with
strong el-el repulsion and forward scattering coupling to phonons.
For a LL with more general el-ph couplings, the kink is still present, but 
the tendency of the EDC to be more peaked for $|E| < \omega_0$ is eliminated
by sufficiently strong el-ph backscattering.
However, in either case, the Fermi liquid analogy cannot be taken too far: 
The basic discrepancy between the
sharpness of the MDCs and the breadth of the EDCs, and
the ``triangular'' confinement of spectral weight in the
$E$-$k$ plane, remain striking aspects of the LL which
differentiate it from the Fermi liquid.

We have also shown examples of measured spectral functions in the cuprates, and drawn attention
to the similarities between them and our
theoretical results.  We believe that the similarities are dramatic.  Exactly why the
measured spectral functions look so much like the
LL coupled to phonons is a deep question,
which we will not explore further, here.
However, at the very least, we feel that this comparison
reinforces the conclusion,
which has certainly been reached \cite{anderson,drorprl,sachdev,vallaNFL} on the basis of a 
variety of other experimental
observations in the cuprates, that these materials are not well described in terms of the
conventional electronic quasiparticles of simple metals.

\begin{acknowledgments}

We wish to thank J. Allen, N. P. Armitage, P. Johnson, T. Valla, S. Sachdev,
G. H. Gweon, E. Arrigoni, D. Orgad, Z.-X. Shen, V. Oganesyan, J. Tranquada, and
A. Lanzara for useful discussions, and X. J. Zhou for supplying
the data in Fig. \ref{contour_experiment}(b).
This work was supported, in part, by the National Science Foundation 
Grant No. DMR 01-10329 (S.A.K.) and by the Department of Energy Contract 
No. DE-FG03-00ER45798 (I.P.B.).

\end{acknowledgments}

\appendix

\section{The spectral function and conductivity of the exactly solvable model}
\label{App}

\begin{figure}
\includegraphics[width=0.495\textwidth]{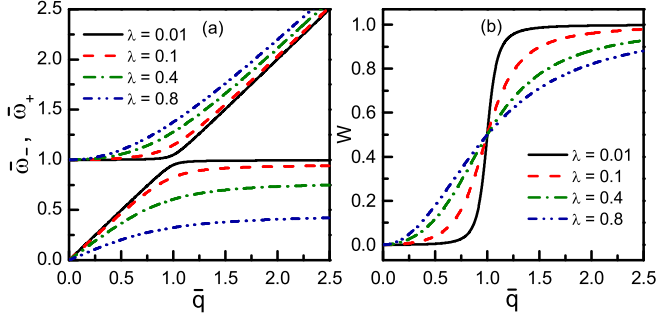}
\caption{\label{spectrum} Panel (a) shows the dispersion of the bosonic modes
$\omega_+$ (upper curves) and $\omega_-$ (lower curves).  (b) shows the spectral weight $W$
of the $\omega_+$ mode in the conductivity.
The notation is
${\bar \omega}_\pm = \omega_\pm/\omega_0$ and ${\bar q} = v_c q/\omega_0$. 
}
\end{figure}

We now derive the spectral function of the model with purely forward scattering
interactions.
We work in the Matsubara representation, in which $\tau = it$ is the imaginary time,
and the partition functional is represented as a path integral over the bosonic fields
$\Pi_c$, $\phi_c$, $\Pi_s$, $\phi_s$, $P$, and $u$.
Since the Lagrangian is quadratic in all of these
fields, we can integrate out
all fields except $\phi_c$ and $\phi_s$.
This results in the Lagrangian ${\cal L} = {\cal L}_c[\varphi_c] + {\cal L}_s[\varphi_s]$ with 
\ba
\label{L}
{\cal L}_c[\varphi_c] = \frac{1}{2 K_c v_c}\left[\omega_n^2 + v_c^2 q^2 - \lambda 
\frac{\omega_0^2 v_c^2 q^2} {\omega_n^2 + \omega_0^2}\right]\left|\varphi_c \right|^2 ,&&\\
{\cal L}_s[\varphi_s] = \frac{1}{2 K_s v_s}\left[\omega_n^2 + v_s^2 q^2\right]\left|\varphi_s 
\right|^2 . \;\;\;\;\;\;\;\;\;\;\;\;\;\;\;\;\;\;\;\;
\ea
Here the bosonic Matsubara frequency is $\omega_n = 2 \pi n/\beta$,
the field $\varphi_\alpha = \varphi_\alpha(q,\omega_n)$ is defined by 
$\phi_\alpha(x,\tau) = (2 \pi \beta L)^{-1} \sum_n \int dq \: e^{-i \omega_n \tau + i q x} 
\varphi_\alpha(q,\omega_n)$,
and $L$ is the length of the system.
The partition functional is then given by the path integral
$Z = \int {\cal D} \varphi_c {\cal D} \varphi_s \, e^{-S_c[\varphi_c] - S_s[\varphi_s]}$
with $S_\alpha [\varphi_\alpha] = (2 \pi \beta L)^{-1} \sum_n \int dq \: {\cal L}_\alpha 
[\varphi_\alpha]$.  
The zero-temperature dispersion relations for the hybridized
charge-phonon bosonic collective modes are thus
\ba
\; \; \nonumber \omega_\pm = \left[\frac{\omega_0^2 + v_c^2 q^2 \pm \sqrt{(\omega_0^2 - v_c^2 
q^2)^2 + 4 \lambda \omega_0^2 v_c^2 q^2}}{2}\right]^{1/2} . \\
\ea
Note that for $\lambda = 0$, $\omega_- = v_c q$ is the long-wavelength density mode of the LL, 
and $\omega_+ = \omega_0$ is the Einstein
phonon mode.  In Fig. \ref{spectrum}(a), we plot $\omega_-$ and $\omega_+$ for various values of 
$\lambda$.

It is a general feature of the bosonization approach that the fermionic correlation functions
are most simply expressed as a function of space and time.
The Matsubara space-time Green's function
was previously computed for a LL
with forward scattering interactions with acoustic
(instead of optical) phonons.\cite{MartinLoss,footnoteappendix}
The analytic structure of Eq. (\ref{L}) differs from the acoustic phonon model according to
$\omega_0 \rightarrow cq$, where $c$ is the velocity of the acoustic phonon.
The initial steps for computing the space-time Green's function involving
functional integration over the fields and Matsubara
frequency summations are the same for our model, except for this
``parameter'' change.  We will therefore not reproduce
these steps here, but rather refer the reader to Ref. \onlinecite{MartinLoss}.
The final momentum integral done in the exponent, however, is different for the two models.

We write the single-particle imaginary-time Green's function for right moving fermions
${\cal G}(x,\tau;\lambda) = - 
\left<T_\tau\Psi_{1,\sigma}(x,\tau)\Psi_{1,\sigma}^\dagger(0,0)\right>$,
where $T_\tau$ is the imaginary time ordering operator, 
as 
\be
{\cal G}(x,\tau;\lambda) = - 
\frac{e^{i k_F x}}{2 \pi a}\exp \left[ - f_c({\bar x}, {\bar \tau}; \lambda) - f_s({\bar x}, 
{\bar \tau})\right]
\ee
(results for left-moving fermions are obtained by changing $x \rightarrow -x$).
Here $f_c$ and $f_s$ are the charge and spin contributions to the exponent, which we will write
in terms of the dimensionless variables ${\bar x}= x \omega_0 /v_c$ and ${\bar \tau}= \omega_0 
\tau$.
To simplify notation, for now we express the result in the limit $T \rightarrow 0$.
For $\tau > 0$ and infinitely large system length, the exact result is
\ba
\label{integral}
\nonumber f_c({\bar x}, {\bar \tau}; \lambda) &=& \int_0^\infty d {\bar q} \: e^{-{\bar a} \bar 
q} \sum_{\nu=\pm} \{ {\cal A}_{\nu} [1 - \textnormal{cos} (\bar{q} \bar{x}) \, 
e^{-\bar{\omega}_\nu \bar{\tau}} ] \\
&-& i \, {\cal B}_{\nu} \, \textnormal{sin} (\bar{q} \bar{x}) \, e^{-\bar{\omega}_\nu \bar{\tau}} 
\} ,
\ea
\be
f_s({\bar x}, {\bar \tau}) \:=\: F({\bar x}, {\bar \tau}; v_s/v_c, K_s, {\bar a}) , 
\;\;\;\;\;\;\;\;\;\;\;\;\;\;\;\;\;\,
\ee
where ${\bar a} = a \omega_0/ v_c$
and we defined the functions
\ba
\label{A_pm} {\cal A}_{\pm}&=& \frac{(K_c^2 + 1)(\bar{\omega}_\pm^2 - 1) + 
\lambda}{4K_c\bar{\omega}_\pm(\bar{\omega}_\pm^2-\bar{\omega}_\mp^2)} \: , \\
{\cal B}_{\pm} &=& \frac{\bar{\omega}_\pm^2 - 1}{2\bar{q}(\bar{\omega}_\pm^2-\bar{\omega}_\mp^2)} 
\: , \\
\label{omega_pm} \bar{\omega}_\pm^2 &=& \frac{\, 1 + \bar{q}^2 \pm \sqrt{(1-\bar{q}^2)^2+4\lambda 
\bar{q}^2} \,}{2} \: ,
\ea
and
\ba
\nonumber F({\bar x}, {\bar \tau}; v, K , a) &=& \frac{1}{8}\left(K+\frac{1}{K}\right) \, \ln 
\left[ \frac{{\bar x}^2 + (a + v{\bar \tau})^2}{a^2} \right]\\
&-& \frac{i}{2} \arctan\left[\frac{{\bar x}}{a + v {\bar \tau}}\right] .
\ea

For $\lambda = 0$, the charge and spin parts are the same
after an appropriate change of parameters:
\be
f_c({\bar x}, {\bar \tau}; 0) = F({\bar x}, {\bar \tau}; 1, K_c, {\bar a}) ,
\ee
yielding the known\cite{pureLLspectralfunction} Green's function of a pure LL
\be
\label{pureLL}
{\cal G}(x,\tau;0) = - \frac{e^{i k_F x}} {2 \pi a} \prod_{\alpha = c,s}H(x, \tau; v_\alpha, 
K_\alpha, a) , 
\ee
where
\ba
\nonumber &&H(x, \tau; v, K, a)\\
&&= \left[ \frac{a^2}{(a + v \tau)^2 + x^2}\right]^{(K-1)^2/8K} \sqrt{\frac{a}{a + v \tau - i x}} 
\, . \; \; \; \; \; \; \;
\ea

For arbitrary values of the parameters,
although it is straightforward to evaluate Eq. (\ref{integral}) numerically, it does
not appear possible to perform the integration analytically.  
We were, however, able to perform it analytically
for the case of arbitrary $\lambda$ but $\bar a \gg 1$:
\ba
\label{deltagreater1} \nonumber {\cal G}(x,\tau;\lambda) = - \frac{e^{i k_F x}} {2 \pi a} H(x, 
\tau; v_c^*, K_c^*, a) \, H(x, \tau; v_s, K_s, a)&&\\ 
{\rm for } \; {\bar a} \gg 1 , \;\;\;\;\;\;\;\;\;\;&& 
\ea
with $v_c^*$ and $K_c^*$ given by Eq. (\ref{vc_star_Kc_star}).
The limit $\bar a \gg 1$ is typically not satisfied in real materials, but
since the spectral function is independent of $\bar a$ for $|{\bar k}|, |{\bar E}| \ll 1/\bar a$,
we can use Eq. (\ref{deltagreater1}) to deduce
the exact behavior of the spectral function
in the limit $|{\bar k}|, |{\bar E}| \ll 1$, regardless of the value of ${\bar a}$:
\ba
{\nonumber}A(k,E;\lambda) \propto A_{LL}(k,E;v_c^*,K_c^*,v_s,K_s) \;\: {\rm for } \; |E| \ll 
\omega_0 ,\\
\ea
where $A_{LL}(k,E;v_c,K_c,v_s,K_s) = A(k,E;0)$ is the spectral function of the pure Luttinger 
liquid.

For the physically interesting case ${\bar a} \ll 1$, we were able to derive
an accurate analytic approximation for Eq. (\ref{integral}), which is the following:
\ba
\label{approx_f} 
&&\nonumber f_c({\bar x}, {\bar \tau}; \lambda) \approx f_c({\bar x}, {\bar \tau}; 0) \; \; \; \; 
\\ 
&&+ \; F({\bar x}, {\bar \tau}; v_c^*/v_c, K_c^*, 1) - F({\bar x}, {\bar \tau}; 1, K_c, 1) , \; 
\; \; \; \; 
\ea
which gives for the Green's function
\ba
\label{final_result}
\nonumber {\cal G}(x,\tau;\lambda) \approx {\cal G}(x,\tau;0)\frac{H(x, \tau; v_c^*, K_c^*, 
v_c/\omega_0)} {H(x, \tau; v_c, K_c, v_c/\omega_0)} && \\
{\rm for } \; {\bar a} \ll 1 ,\;\;\;\;\;\;\;\;\;&& 
\ea
with $v_c^*$ and $K_c^*$ given again by Eq. (\ref{vc_star_Kc_star}).
The $T \neq 0$ single-hole real-time Green's function is written,
using this approximation, in Eq. (\ref{main_result}).

\begin{figure}
\includegraphics[width=0.49\textwidth]{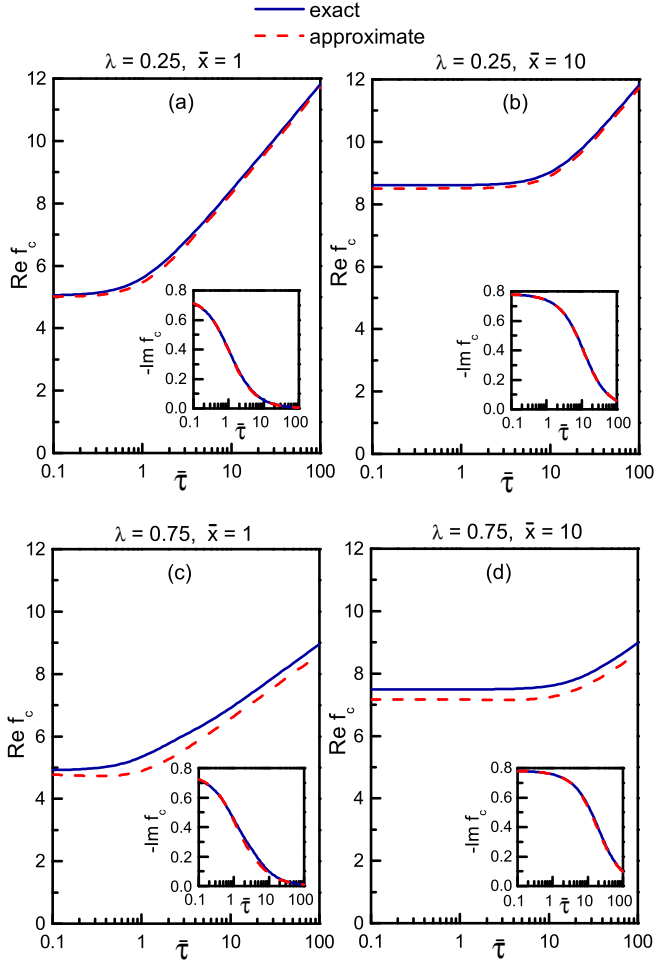}
\caption{\label{exact_vs_approx} Comparison of the exact
Green's function charge exponent $f_c({\bar x},{\bar \tau};\lambda)$,
obtained by evaluating 
Eq. (\ref{integral}) numerically (solid lines), with the approximate analytic
expression in Eq. (\ref{approx_f}) (dashed lines).
Panels (a) and (b) are for $\lambda = 0.25$;
(c) and (d) are for $\lambda = 0.75$.
For all panels, $\gamma_c = 0.6$ and ${\bar a} = 0.05$.
Panels (a) and (c) show the ${\bar \tau}$ dependence for ${\bar x = 1}$;
(b) and (d) are for ${\bar x = 10}$.  The main plots show the real part of $f_c$ and
the insets show the imaginary part.  For the imaginary part, the dashed and solid lines
overlap.
}
\end{figure}

We demonstrate the accuracy of this analytic approximation by comparing
Eq. (\ref{approx_f})
with the exact result, obtained by performing the integration
in Eq. (\ref{integral}) numerically.
From Figs. \ref{exact_vs_approx}(c) and
\ref{exact_vs_approx}(d) we see that for $\lambda = 0.75$
and $\gamma_c = 0.6$,
the approximation gives 
within $9\%$ of the exact result for the real part of $f_c({\bar x}, {\bar \tau}; \lambda)$.
For $\lambda = 0.25$ [Figs. \ref{exact_vs_approx}(a) and \ref{exact_vs_approx}(b)],
the error is less than $3\%$.  The agreement with the imaginary part is similarly excellent.

We have also computed the frequency- and momentum-dependent
conductivity for the LL with forward scattering 
off phonons.  The exact result for the real part of the conductivity at $T = 0$ is
\ba
\nonumber \sigma_1(q,\omega; \lambda) &=& \nonumber e^2 K_c v_c \sum_{\nu = \pm 1}[W 
\delta(\omega - \nu \omega_+) \\
&+& \, (1-W)\delta(\omega - \nu \omega_-)] . \; \; \;
\label{sigma}
\ea
Here $e$ is the charge of the electron, $\delta$ is the delta function, and the spectral
weight of the $\omega_+$ mode is
\be
W = \frac{\omega_+^2-\omega_0^2}{\omega_+^2-\omega_-^2} \: ,
\ee
which depends only on $v_c q/\omega_0$ and $\lambda$.  We plot $W$ 
in Fig. \ref{spectrum}(b).
Note that the conductivity at $q = 0$ (optical conductivity)
remains unchanged by forward scattering off phonons:
\be
\sigma_1(0,\omega;\lambda) = \sigma_1(0,\omega;0) = 2 e^2 K_c v_c \delta(\omega) ,
\ee  
yet this same interaction produces dramatic changes to the spectral function.

\section{RG treatment of general electron-phonon coupling}
\label{AppB}

In the case in which the el-el couplings and the
backscattering el-ph coupling $\lambda_1$ [Eq. (\ref{barecouplings})]
are weak,
although of arbitrary relative strength, the phonon-induced
renormalization of the LL parameters
can be computed using a two-step
perturbative (one-loop) RG scheme.\cite{ZKL,Voit,ian}
Since, for $E_F \gg \omega_0$, the coupling
$\lambda_1$ is strongly renormalized,
and the RG flows of $\lambda_1$
are modified by the direct el-el interactions,
the resulting expressions for the renormalizations of $v_c$ and $K_c$
are more complicated than in Eq. (\ref{vc_star_Kc_star}).
However, they have recently been
analyzed in detail
by one of us,\cite{ian} and will be summarized below.

We employ standard notation for 
the important short-range el-el interaction parameters of the incommensurate 1DEG: $g_1$ 
(backscattering), $g_2$
(forward scattering on both left- and right-moving branches), and $g_4$ (forward scattering on 
only one branch).
The Hamiltonian density for the el-el interaction portion is then
\ba
\nonumber {\cal H}_{\rm el-el} &=& g_1 \sum_{\sigma,\sigma^\prime = \pm1} \Psi_{1,\sigma}^\dagger 
\Psi_{-1,\sigma^\prime}^\dagger \Psi_{1,\sigma^\prime} \Psi_{-1,\sigma} \\
\nonumber &+& g_2 \sum_{\sigma,\sigma^\prime = \pm1} \Psi_{1,\sigma}^\dagger 
\Psi_{-1,\sigma^\prime}^\dagger \Psi_{-1,\sigma^\prime} \Psi_{1,\sigma} \\
&+& g_4 \sum_{\eta,\sigma = \pm1} \Psi_{\eta,\sigma}^\dagger \Psi_{\eta,-\sigma}^\dagger 
\Psi_{\eta,-\sigma} \Psi_{\eta,\sigma} .
\label{H_el-el}
\ea
For the extended Hubbard model in the continuum (weak-coupling) limit,
these parameters are $g_1 = U - 2V^\prime$, $g_2 = U + 2V^\prime$, and $g_4 = U/2 + 2V^\prime$,
where $U$ is the on-site interaction,
$V^\prime = -V \cos(2 k_F)$, and $V$ is the nearest neighbor interaction
(near halffilling, $V^\prime \approx V$).

For an incommensurate 1DEG in the presence of a general el-ph coupling, the effective charge 
parameters
at low energy are
\ba
\label{vc_star_with_backscattering} v_c^* &=& v_F\sqrt{(1 + g_4^{\rm tot})^2 - (g_c^{\rm 
tot}/2)^2} \, , \; \; \\
\label{Kc_star_with_backscattering} K_c^* &=& \sqrt{\frac{1 + g_4^{\rm tot} + g_c^{\rm tot}/2}{1 
+ g_4^{\rm tot} - g_c^{\rm tot}/2}} \, , \; \;
\ea
where we defined
\ba
\label{g1} g_c^{\rm tot} &=& \bar{g}_1 - 2\bar{g}_2 - (\lambda_1^*  - 2\lambda_2) ,\\
g_4^{\rm tot} &=& \bar{g}_4 - \lambda_2 ,
\ea
and ${\bar g_i} = g_i/\pi v_F$.  Here the renormalized el-ph backscattering coupling is denoted
by $\lambda_1^*$.
For the case $E_F \gg \omega_0$ and ${\bar g}_1, {\bar g}_2, \lambda_1 \ll 1$, it is given 
by\cite{ian}
\be
\lambda_1^* = \frac{\lambda_1 h(l_0)}{1 - \lambda_1 \int_0^{l_0} dx \, h(x)} \: ,
\label{lambda_star}
\ee
where
\be
h(x) = \frac {\exp\left[-({\bar g}_1 - 2{\bar g}_2) x/2 \, \right]} {(1 + \bar{g}_1 x)^{3/2}} \,,
\ee
$l_0 \equiv \ln(E_F/\omega_0)$, and $\lambda_1$ and $\lambda_2$ are the bare el-ph couplings 
defined in Eq. (\ref{barecouplings}).

In the weak-coupling limit and for $E_F \gg \omega_0$,
the total effective backscattering interaction
after integrating out degrees of freedom from $E_F$ to $\omega_0$
is given by
\be
g_1^{\rm tot} = \frac{{\bar g}_1}{1 + {\bar g}_1 l_0} - \lambda_1^* \, .
\ee
In order for the Luttinger liquid description to be valid,
it is required that $g_1^{\rm tot}$
is repulsive ($g_1^{\rm tot} > 0$).
In this case,
the effective spin parameters are $K_s^* = 1$ and $v_s^* = v_F(1 - {\bar g}_4)$.
However, if $\lambda_1^*$ is large enough to cause $g_1^{\rm tot} < 0$, 
the RG flows carry $K_s^*$ and $v_s^*$ to zero at low energies, 
signaling the emergence of a spin gap.
In this case, the system is described as a Luther-Emery liquid\cite{LEL} phase.
We refer the reader to Ref. \onlinecite{ian} for a detailed study
of the phase diagram of the 1DEG coupled to phonons.  Since the spectral function
computed in the present paper is not applicable to a Luther-Emery liquid,
here we restrict our attention to the case in which there is no spin gap.

By combining the above expressions for
$v_c^*$ and $K_c^*$ with the expressions for $v_c$ and $K_c$
[given by Eqs. (\ref{vc_star_with_backscattering})
and (\ref{Kc_star_with_backscattering}) with $\lambda_1^* = \lambda_2 = 0$],
one can rewrite the results in a form that shows more clearly how 
the relation $v_c^*/v_c = K_c/K_c^* = \sqrt{1 - \lambda}$
is modified by the presence of el-ph backscattering--see Eqs. (\ref{new_vc}) and (\ref{new_Kc}).
There, the effects of el-ph backscattering are contained
in the parameter $\Lambda$, defined in Eq. (\ref{cap_lambda}).


\begin{references}

\bibitem{shen1}A. Lanzara, P. V. Bogdanov, X. J. Zhou, S. A. Kellar, D. L. Feng,
E. D. Lu, T. Yoshida, H. Eisaki, A. Fujimori, K. Kishio, J.-I. Shimoyama, T. Noda, S. Uchida, Z. 
Hussain, and Z.-X. Shen,
Nature (London) {\bf 412}, 510 (2001).
\bibitem{shen2}X. J. Zhou, T. Yoshida, A. Lanzara, P. V. Bogdanov, S. A. Kellar, K. M. Shen, W. 
L. Yang, F. Ronning, T. Sasagawa, T. Kakeshita, T. Noda, H. Eisaki, S. Uchida, C. T. Lin, F. 
Zhou, J. W. Xiong, W. X. Ti, Z. X. Zhao, A. Fujimori, Z. Hussain, and Z.-X. Shen,
Nature (London) {\bf 423}, 398 (2003).
\bibitem{lanzara}G.-H. Gweon, T. Sasagawa, S. Y. Zhou, J. Graf, H. Takagi, D.-H. Lee, and A. 
Lanzara, Nature (London) {\bf 430}, 187 (2004).
\bibitem{anderson}P. W. Anderson, {\it The Theory of Superconductivity in the High-$T_c$ 
Cuprates}
(Princeton University Press, Princeton, NJ, 1997).
\bibitem{drorprl}D. Orgad, S. A. Kivelson, E. W. Carlson,
V. J. Emery, X. J. Zhou, and Z.-X. Shen, Phys. Rev. Lett. {\bf 86}, 4362 (2001).
\bibitem{sachdev}M. Vojta, Y. Zhang, and S. Sachdev, Int. J. Mod. Phys. B {\bf 14}, 3719 (2000).
\bibitem{vallaNFL}T. Valla, A. V. Fedorov, P. D. Johnson, B. O. Wells, S. L. Hulbert, Q. Li, G. 
D. Gu,
and N. Koshizuka, Science {\bf 285}, 2110 (1999).
\bibitem{meden}V. Meden, K. Sch\"{o}nhammer, and O. Gunnarsson, Phys. Rev. B {\bf 50}, 11179 
(1994).
\bibitem{bourbonnais}C. Bourbonnais,
in {\it High Magnetic Fields: Applications in Condensed Matter Physics and Spectroscopy},
edited by C. Berthier, L. P. Levy, and G. Martinez
(Springer-Verlag, Berlin, 2002), p. 235.
\bibitem{nanotubes}R. Egger, A. Bachtold, M. Fuhrer, M. Bockrath, D. Cobden, and P. McEuen,
in {\it Interacting Electrons in Nanostructures}, edited by R. Haug and H. Schoeller
(Springer-Verlag, Berlin, 2001), p. 125. 
\bibitem{gweon}G.-H. Gweon, J. D. Denlinger, J. W. Allen, R. Claessen, C. G. Olson, H. 
H\"{o}chst,
J. Marcus, C. Schlenker, L. F. Schneemeyer, and G. Gweon, J. Electron Spectrosc. Relat. Phenom. 
{\bf 117}, 481 (2001).
\bibitem{allen}J. W. Allen, Solid State Commun. {\bf 123}, 469 (2002).
\bibitem{GweonAllen}G.-H. Gweon, J. W. Allen, and J. D. Denlinger, Phys. Rev. B {\bf 68}, 195117 
(2003).
\bibitem{stripereview}For a recent review of the experimental evidence regarding the 
existence of local stripe correlations in the high-temperature superconductors,
see S. A. Kivelson, I. P. Bindloss, E. Fradkin, V. Oganesyan, J. M. Tranquada, A. Kapitulnik, and 
C. Howald,  
Rev. Mod. Phys. {\bf 75}, 1201 (2003).
\bibitem{footnotestripes}In the present paper, we compare
our results to ARPES measurements
in the cuprates in the nodal
direction ${\bf k}$ = (0,0)-($\pi$,$\pi$).
This is done without microscopic justification.
However, it is worth mentioning that in Refs. \onlinecite{mats} and \onlinecite{mats2},
the band structure of a disordered array 
of stripes is computed, and is found to have a nodal component.
This implies that the 1D physics along the stripes is being mirrored in this 
direction.
We also mention that, according to Ref. \onlinecite{luther},
effective 1D physics can emerge
in a 2D system with straight Fermi segments and strong interactions.
\bibitem{marsiglio}For a recent review, see F. Marsiglio and J. P. Carbotte, cond-mat/0106143 
(unpublished).
\bibitem{footnotebroadening}We do not mean to imply that phonons produce no broadening
at high binding energies in the LL.  They indeed do, in a way analogous
to the Fermi liquid, but for $K_c \ll 1$ this effect is negligible
compared to the huge amount of broadening already present in the EDCs
from direct el-el interactions.
In contrast, the narrowing effect
for $|E| < \omega_0$, due to an increase in the effective $K_c$, is dramatic in the LL
but not present in the Fermi liquid.
\bibitem{emery}V. J. Emery, in {\it Highly Conducting
One-Dimensional Solids}, edited by J. T. Devreese, R. P. Evrard, and
V. E. van Doren (Plenum, New York, 1979), p. 247.
\bibitem{fradkinbook}E. Fradkin, {\it Field Theories of Condensed Matter Systems} 
(Addison-Wesley, Reading, MA, 1991).
\bibitem{voit}J. Voit, Rep. Prog. Phys. {\bf 58}, 977 (1995).
\bibitem{tsvelik}A. O. Gogolin, A. A. Nersesyan, and A. M. Tsvelik,
{\it Bosonization and Strongly Correlated Systems} (Cambridge University Press, Cambridge, U.K., 
1998).
\bibitem{giamarchi}T. Giamarchi, {\it Quantum Physics in One Dimension} (Oxford University Press, 
Oxford, 2004).
\bibitem{MartinLossPRB}D. Loss and T. Martin, Phys. Rev. B {\bf 50}, 12160 (1994).
\bibitem{pureLLspectralfunction}For the
spectral function in the absence of electron-phonon coupling, see A. Luther and I. Peschel, Phys. 
Rev. B. {\bf 9}, 2911 (1974);
V. J. Emery, in {\it Highly Conducting
One-Dimensional Solids}
(Ref. \onlinecite{emery}); V. Meden and K. Sch\"{o}nhammer, Phys. Rev. B {\bf 46}, 15753 (1992);
J. Voit, Phys. Rev. B, {\bf 47}, 6740 (1993); D. Orgad, Philos. Mag. B {\bf 81}, 375 (2001); T. 
Giamarchi, {\it Quantum Physics in One Dimension} (Ref. \onlinecite{giamarchi}).  
\bibitem{footnotesinglehole}The single-hole spectral function
for $k$ near $-k_F$ is given by $A(-k,E)$.
The {\it single-particle} spectral function for $k$ near $\pm k_F$
is $A(\mp k,-E)$.  The full
spectral function for $k$ near $\pm k_F$ is given by
$A(\pm k,E) + A(\mp k,-E)$. 
\bibitem{footnoteunits}We plot the spectral function in arbitrary units and
therefore ignore the dependence of the numerical prefactor on $\omega_0$.
The cutoff ${\bar a}$
is also not considered a free parameter, since
the spectral function is independent of its value so long as
$1/{\bar a} \gg |{\bar k}|, |{\bar E}|$.
For all plots, we used ${\bar a} = 0.05$.
\bibitem{LEL}A. Luther and V. J. Emery, Phys. Rev. Lett. {\bf 33}, 589 (1974).
\bibitem{ian}I. P. Bindloss, cond-mat/0404154 (unpublished). 
\bibitem{footnoteinteresting}It is interesting to note
that the instability at large $\lambda$
mentioned in Sec. \ref{exact_model} survives the addition 
of el-ph backscattering, but is moved
from $\lambda = 1$ to $\lambda = 1 + K_c^2\Lambda$.
\bibitem{shen3}P. V. Bogdanov, A. Lanzara, S. A. Kellar, X. J. Zhou, E. D. Lu, W. J. Zheng,
G. Gu,  J.-I. Shimoyama, K. Kishio, H. Ikeda, R. Yoshizaki, Z. Hussain, and Z. X. Shen,
Phys. Rev. Lett. {\bf 85}, 2581 (2000).
\bibitem{CampuzanoKink}A. Kaminski, M. Randeria, J. C. Campuzano, M. R. Norman, H. Fretwell,
J. Mesot, T. Sato, T. Takahashi, and K. Kadowaki, 
Phys. Rev. Lett. {\bf 86}, 1070 (2001).  
\bibitem{johnson}P. D. Johnson, T. Valla, A. V. Fedorov, Z. Yusof, B. O. Wells, Q. Li, A. R. 
Moodenbaugh, G. D. Gu, N. Koshizuka,
C. Kendziora, Sha Jian, and D. G. Hinks, Phys. Rev. Lett. {\bf 87}, 177007 (2001).
\bibitem{footnotebackground}Some authors have suggested that a part of the measured $A({\bf 
k},E)$ is due to 
extrinsic processes which also produce a ``background'' signal at wave 
vectors far outside the Fermi surface.  If such a background is 
subtracted from the measured signals, it reduces, but does not 
eliminate the long high energy tails seen in the EDCs.
We believe that good fits could be obtained to the data in this 
case, too, although likely with somewhat smaller values of $\gamma_c$.
\bibitem{Campuzano}A. Kaminski, J. Mesot, H. Fretwell, J. C. Campuzano, M. R. Norman, M. 
Randeria, H. Ding, T. Sato, T. Takahashi, T. Mochiku, K. Kadowaki, and H. Hoechst,
Phys. Rev. Lett. {\bf 84}, 1788 (2000).
\bibitem{footnotekink}For 2D or 3D systems
that contain a kink due to el-ph coupling,
the dispersion at high binding energies is seen to extrapolate to the origin
in both theory (Ref. \onlinecite{Verga}) and experiment (Ref. \onlinecite{valla}).
\bibitem{LSCO}T. Yoshida, X. J. Zhou, T. Sasagawa, W. L. Yang, P. V. Bogdanov, A. Lanzara, Z.
Hussain, T. Mizokawa, A. Fujimori, H. Eisaki, Z.-X. Shen, T. Kakeshita, and S. Uchida,
Phys. Rev. Lett. {\bf 91}, 027001 (2003).
\bibitem{Verga}S. Verga, A. Knigavko, and F. Marsiglio, Phys. Rev. B {\bf 67}, 054503 (2003).
\bibitem{Schachinger}E. Schachinger, J. J. Tu, and J. P. Carbotte, Phys. Rev. B {\bf 67}, 214508 
(2003). 
\bibitem{MartinLoss}T. Martin and D. Loss, Int. J. Mod. Phys. B {\bf 9}, 495 (1995).
\bibitem{footnoteappendix}The single-particle density of states (momentum integrated spectral 
function)
for a different model of a LL coupled to acoustic phonons was investigated in Ref. 
\onlinecite{papa}.
\bibitem{ZKL}G. T. Zimanyi, S. A. Kivelson, and A. Luther, Phys. Rev. Lett. {\bf 60}, 2089 
(1988).
\bibitem{Voit}J. Voit and H. J. Schulz, Phys. Rev. B {\bf 37}, 10068 (1988).
\bibitem{mats}M. Granath, V. Oganesyan, D. Orgad, and S. A. Kivelson, Phys. Rev. B {\bf 65}, 
184501 (2002).
\bibitem{mats2}M. Granath, Phys. Rev. B {\bf 69}, 214433 (2004).
\bibitem{luther}A. Luther, Phys. Rev. B {\bf 50}, 11446 (1994).
\bibitem{valla}T. Valla, A. V. Fedorov, P. D. Johnson, and S. L. Hulbert, Phys. Rev. Lett. {\bf 
83}, 2085 (1999).
\bibitem{papa}E. Papa and A. M. Tsvelik, cond-mat/0004007 (unpublished).
 
\end{references}
\end{document}